\documentclass[runningheads]{llncs}
\usepackage[T1]{fontenc}
\usepackage[colorlinks,linkcolor=green,anchorcolor=green,citecolor=green]{hyperref}
\usepackage{graphicx}
\usepackage{multirow}
\usepackage{amsfonts,amssymb} 
\usepackage[misc]{ifsym}
\usepackage[ruled]{algorithm2e}
\begin{document}
\title{Unsupervised Decomposition Networks for Bias Field Correction in MR Image}%
%\thanks{Supported by organization x.}}
%
\titlerunning{Unsupervised Decomposition Networks}
\author{Dong Liang\inst{1}, Xingyu Qiu\inst{1}, Kuanquan Wang$^{(\textrm{\Letter})}$\inst{1}, Gongning Luo$^{(\textrm{\Letter})}$\inst{1}, Wei Wang\inst{1}, Yashu Liu\inst{1}}%
\authorrunning{D. Liang et al.}
\institute{Harbin Institute of Technology, Harbin, China\\
\email{\{wangkq, luogongning\}@hit.edu.cn}
}
\maketitle              % typeset the header of the contribution
\begin{abstract}
Bias field, which is caused by imperfect MR devices or imaged objects, introduces intensity inhomogeneity into MR images and degrades the performance of MR image analysis methods. Many retrospective algorithms were developed to facilitate the bias correction, to which the deep learning-based methods outperformed. However, in the training phase, the supervised deep learning-based methods heavily rely on the synthesized bias field. As the formation of the bias field is extremely complex, it is difficult to mimic the true physical property of MR images by synthesized data. While bias field correction and image segmentation are strongly related, the segmentation map is precisely obtained by decoupling the bias field from the original MR image, and the bias value is indicated by the segmentation map in reverse. Thus, we proposed novel unsupervised decomposition networks that are trained only with biased data to obtain the bias-free MR images. Networks are made up of: a segmentation part to predict the probability of every pixel belonging to each class, and an estimation part to calculate the bias field, which are optimized alternately. Furthermore, loss functions based on the combination of fuzzy clustering and the multiplicative bias field are also devised. The proposed loss functions introduce the smoothness of bias field and construct the soft relationships among different classes under intra-consistency constraints. Extensive experiments demonstrate that the proposed method can accurately estimate bias fields and produce better bias correction results. The code is available on the link \url{https://github.com/LeongDong/Bias-Decomposition-Networks}.

\end{abstract}
\section{Introduction}
Magnetic resonance imaging (MRI) techniques provide abundant anatomical details, which are critical to precise diagnosis and prognosis. The bias field is a common phenomenon in MR image created by imperfect MR devices or imaged objects. It brings artifactual signal inhomogeneity that intensity within the same tissue varies smoothly across the MR image, which could degrade the subsequent quantitative analysis tasks.
\begin{figure}
\includegraphics[width=9cm]{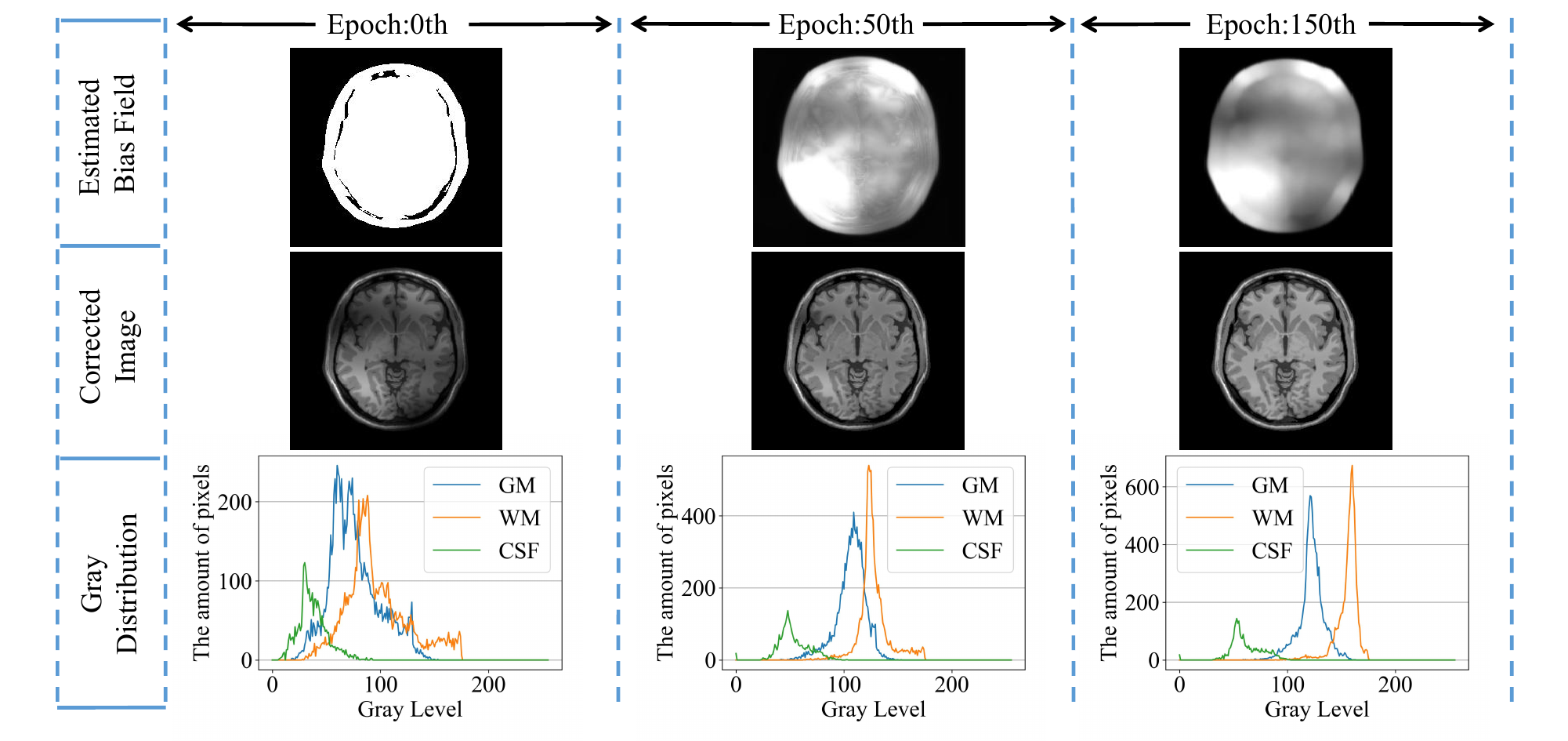}
\centering
\caption{The training progress of the proposed model to remove the bias field from the original bias-influenced MR image. In $0th$ epoch, the corrected image is the original input image, and the estimated bias field is initialized into a binary mask by thresholding. (CSF=Cerebro-Spinal Fluid,GM=Gray Matter,WM=White Matter.)} \label{fig1}
\end{figure}
To deal with this problem, many bias correction methods have been proposed, which are divided into prospective methods and retrospective methods~\cite{ref_article1}. Compared with the prospective methods that rectify the bias in the process of signal acquisition, the retrospective methods are executed based on the image information and can remove the bias field induced by both MR devices and imaged objects~\cite{ref_article2}. The popular retrospective methods are further classified into histogram-~\cite{ref_article3}\cite{ref_article4}, segmentation-~\cite{ref_proc1,ref_article5,ref_article6,ref_article7,ref_article8,ref_article9,ref_article10} and deep learning-based methods\cite{ref_proc3}\cite{ref_article11}. As the bias field varies smoothly in low frequency, Sled et al.~\cite{ref_article3} proposed a nonparametric nonuniformity normalization ($N3$) method for bias field correction by iteratively maximizing high frequency within the tissue. In order to improve the $N3$ method, Tustison et al.~\cite{ref_article4} selected control points from $N$ neighborhoods and adopted a multiresolution approximation strategy to obtain faster and effective convergency performance. Unfortunately, when the range of the bias value is large enough, the method named $N4$ tends to produce  unrealistic results. Ahmed et al.~\cite{ref_article8} pointed out that the segmentation map can be used to compensate for the bias field. However, this segmentation-based method neglects the smoothness constraint of bias field, which could incorrectly cause a single bin result. Li et al.~\cite{ref_proc1} introduced level-set for segmentation tasks. However, this method relies on manual initialization and cannot be executed automatically. 
  
  Nowadays, deep learning-based methods have achieved remarkable performance in automatic medical image analysis. Goldfryd et al.~\cite{ref_proc3} devised a semi-supervised compound framework for estimating the bias-free MR image and bias field. Similarly, Chen et al.~\cite{ref_article11} created a generative network to generate bias field and a discriminative network for supervision. However, the supervised deep learning-based methods suffer from the shortage of bias field ground truth. Although synthesized data is applied in supervision, it hardly reflects the physical properties of true bias-influenced MR image.
  
  To solve the above-mentioned problems, we proposed unsupervised decomposition networks to fully automatically estimate the bias field. The proposed decomposition networks include segmentation part and bias estimation part for alternate optimization. The networks gradually recover the bias-influenced image to a bias-free image during training, as shown in Fig.~\ref{fig1}. The main contributions of our research can be concluded as follows: a) We proposed, to our best knowledge, the first unsupervised deep learning based model for bias field estimation, which is trained on bias-influenced MR images; b) We designed the probability map and the bias field reconstruction loss functions based on the combination of fuzzy clustering and properties of bias field. The loss functions introduce the smoothness and multiplicative characteristics for bias field and constructs the soft relationships among classes under the intra-consistency constraints, which helps the model effectively learn the inherent data structures; c) Extensive experiments executed on both synthesized and real MRI datasets demonstrate the effectiveness of our proposed method to create accurate bias corrected results.
\section{Method}
\subsubsection{Principles and framework}
Based on the principles of bias formation, the bias fields can be precisely decomposed from MR images. In most previous studies, the bias field was described in a simple multiplicative form as:
\begin{equation}
I(r)=i(r)b(r)+n(r), \forall r\in\Omega\label{multi-bias}
\end{equation}
where $r$ denotes the location in the image field $\Omega$, $I(\cdot)$ is the acquired MR image, $i(\cdot)$ is the corresponding bias-free MR image without bias field, $b(\cdot)$ represents the bias field, and $n(\cdot)$ is the noise, which could be estimated by quasi-Gaussian functions. For a bias-free MR image, pixels within the same tissue are assumed to possess the similar gray distribution. Thus, the gray value of each pixel can be approximated by corresponding class center in reverse. Inspired by fuzzy c-means (FCM) algorithm, we constructed an energy function based on Eq.\ref{multi-bias} and minimized it to estimate the appropriate bias field, which is formulated as:
\begin{equation}
\min\ E_{u,c,b}=\sum_{i=1}^{N_{c}}\int_{\Omega}u_{i}^{p}(r)||I(r)-b(r)c_{i}||^{2}dr, \qquad \textrm{subject\ to} \sum_{i=1}^{N_{c}}u_{i}(r)=1\label{Energy}
\end{equation}
where $N_{c}$ is the number of classes with different gray distributions, $u_{i}(r)$ represents the probability that the $r$th pixel is classified into the $i$th category, and $c_{i}$ is $i$th class center. The parameter $p$ determines the fuzziness of the classification process. Also, for a certain pixel, the sum of probabilities belonging to each class is restricted to $1$.\\
\begin{figure}[h]
\includegraphics[width=9.5cm]{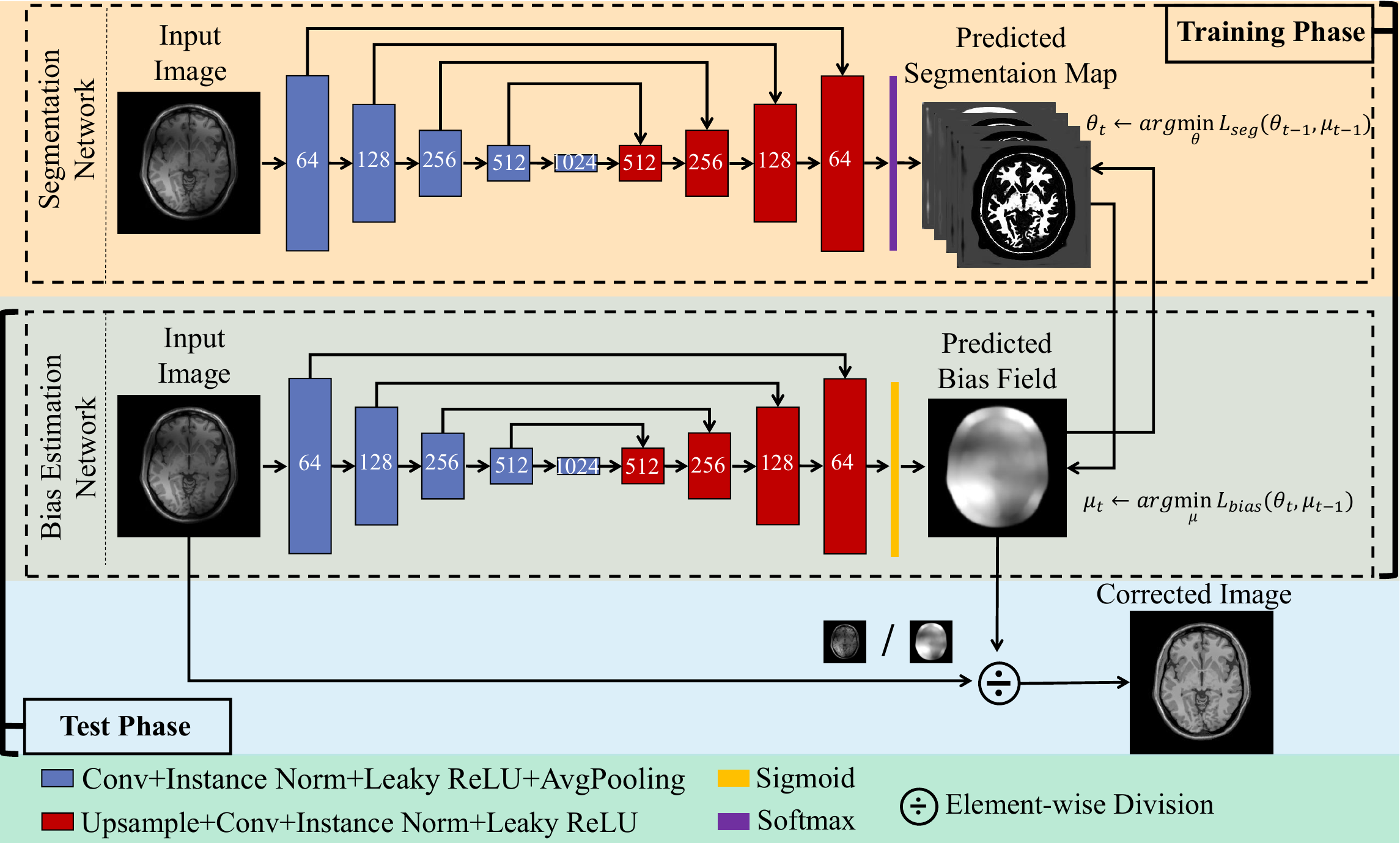}
\centering
\caption{Overview of proposed unsupervised bias decomposition networks.} \label{fig2}
\end{figure}
\quad The relationship between the energy $E_{u,c,b}$ minimization and bias field $b(r)$ estimation is that the homogeneity within each class could be improved through accurate bias field correction, while the energy would be minimized by ensuring intra-class consistency. Due to the differentiable property of energy function $E_{u,c,b}$, the local extrema, as a sufficient but not neccessary condition, are acquired by calculating the first derivates equaling zero on variables $u_{i}(r)$, $c_{i}$ and $b(r)$, respectively. The computed results are illustrated as follows:
\begin{equation}
u_{i}(r)=\frac{1}{\sum_{j=1}^{N_{c}}(\frac{||I(r)-b(r)c_{i}||^{2}}{||I(r)-b(r)c_{j}||^{2}})^{\frac{1}{p-1}}}\label{probability}
\end{equation}
\begin{equation}
c_{i}=\frac{\int_{\Omega}b(r)I(r)u_{i}^{p}(r)dr}{\int_{\Omega}b^{2}(r)u_{i}^{p}(r)dr}\label{center}
\end{equation}
To keep the smoothness constraint on the bias field, we introduce a simple but effective Gaussian filter in the process of bias field calculation as follows:
\begin{equation}
b(r)=\frac{K*\sum_{i=1}^{N_{c}}c_{i}I(r)u_{i}^{p}(r)}{K*\sum_{i=1}^{N_{c}}c_{i}^{2}u_{i}^{p}(r)}\label{bias}
\end{equation}
where $K$ is an $n\times n$ Gaussian filter and $*$ is convolution operation. Variables $c_{i}$, $u_{i}(r)$ and $b(r)$ are alternately optimized by an expectation-maximization scheme. To simplify the iterative process, we choose the class center $c_{i}$ as an intermediate variable and design novel bias decomposition networks to estimate the variables $u_{i}(r)$ and $b(r)$. As shown in Fig.~\ref{fig2}, the proposed model is composed of two parts: a segmentation network for predicting the probability of every pixel belonging to each class, and a bias estimation network for calculating the bias field. Both of the two networks are based on the same structure as basic U-Net in ~\cite{ref_proc2}. The only difference is that the segmentation network adopts the Softmax function to predict the probability map, and the bias estimation network applies the Sigmoid function to restrict the range of bias field in the last output layer.
\subsubsection{Loss functions and training strategy}Two reconstruction loss functions (Eq.\ref{lossbias} \& Eq.\ref{lossseg}) are designed to ensure the smooth and multiplicative characteristics of bias field $b$ and the intra-class consistent constraints for the probability map $u$. The training details are described as Algorithm~\ref{alg:algorithm1}. The outputs of the segmentation network and bias estimation network are firstly copied into two new variables $u_{d}$ and $b_{d}$. Then, $u_{d}$ and $b_{d}$ are detached from the computational graph without gradient back-propagation. After that, the original image $I$, $u_{d}$ and $b_{d}$ are used to reconstruct the rectified probability map $u$ and bias field $b$ by Eq.~\ref{probability}-\ref{bias}. Finally, the reconstructed results, regarded as the ground truth, are applied in supervision on network training. The whole process is repeated until $\frac{1}{||\Omega||}\int_{\Omega}||b_{t}(r)-b_{t-1}(r)||^2dr<\epsilon$, where $\epsilon$ is a small number.
\begin{algorithm}[H]
\caption{Training on bias decomposition networks.}
\label{alg:algorithm1}
\LinesNumbered
\KwIn{original image $I$, fuzziness $p$.}
\KwOut{Updated parameters $\theta$ and $\mu$ of bias decomposition networks.}
Input $I$ to obtain the output $u^{pred}$ for segmentation network and $b^{pred}$ for bias estimation network. Copy $u^{pred}$ to $u_{d}$, $b^{pred}$ to $b_{d}$, and close the gradient back-propagation of both $u_{d}$ and $b_{d}$;\\
Calculate class center $c$ using $I$, $p$, $u_{d}$ and $b_{d}$ by Eq.\ref{center}.\\
Calculate rectified probability map $u$ using $I$, $p$, $c$ and $b_{d}$ by Eq.\ref{probability}.\\
Construct loss function for training segmentation network as:
\begin{equation}
Loss_{seg}=\sum_{i=1}^{N_{c}}\int_{\Omega}||u_{i}(r)-u_{i}^{pred}(r)||^{2}dr\label{lossseg}
\end{equation}\\
Update the segmentation network's parameters $\theta$ as:
\begin{equation}
\theta_{t}\leftarrow\arg\min\limits_{\theta}{Loss_{seg}(\theta_{t-1},\mu_{t-1})}
\end{equation}
where $t$ is the alternation index, $\leftarrow$ represents message passing and $\mu$ is the parameter of bias estimation network.\\ 
Update $u_{d}$ and $c$. Calculate rectified bias field $b$ using $I$, $p$, $c$, $u_{d}$ by Eq.\ref{bias}.\\
Construct loss function for training bias estimation network as:
\begin{equation}
Loss_{bias}=\int_{\Omega}||b(r)-b^{pred}(r)||^{2}dr
\label{lossbias}
\end{equation}\\
Update the bias estimation network's parameters $\mu$ as:
\begin{equation}
\mu_{t}\leftarrow\arg\min\limits_{\mu}{Loss_{bias}(\theta_{t},\mu_{t-1})}
\end{equation}\\
Repeate 1-8, until convergency.
\end{algorithm}
\section{Experiment}
\subsubsection{Materials and Evaluation Metrics}
The proposed method is validated on the BrainWeb Dataset\cite{ref_article13} and OAI-ZIB Dataset\cite{ref_article12}. The BrainWeb dataset includes T1-, T2-sequences. As all sequences are simulated without bias field, the BrainWeb dataset is used as the benchmark \cite{ref_proc3}. The OAI-ZIB dataset includes real MR data in DESS sequences. In our experiment, we randomly selected 10 volumes for training and another 10 volumes for testing from the OAI-ZIB dataset. To evaluate bias correction methods, we adopt \textbf{coeffcient of variation} to quantify the intensity inhomogeneity within each tissue. Meanwhile, we use \textbf{Structural Similarity} (SSIM) and \textbf{Peak Signal to Noise Ratio} (PSNR) to evaluate the similarity between bias-free image and bias-corrected image.
\subsubsection{Implementation Details}
For the training of networks, the batch size is set to $1$ and the initial learning rate is 0.001 adjusted by Adam optimizer with 0.5 decay rate for every 100 epochs. The training phase will be stopped when the mean square error on bias fields within 2 epochs is less than $10^{-6}$. The kernel size of Gaussian filter is 5. To ensure the convergency of conventional methods, the iteration numbers are set to 200. The fuzziness $p$ and the number of classes $N_{c}$ are set to 2 and 4, respectively. Besides, we use Legendre polynomials \cite{ref_article11} and trigonometric functions to create synthesized bias fields. The simulated 2D bias field is described as follows: 
\begin{equation}
b(x,y)=\sum_{i=0}^{L_{l}}\sum_{j=0}^{L_{l}-i}w_{i,j}P_{i}(x)P_{j}(y)+\sum_{l=0}^{L_{t}}\sum_{k=0}^{l}w_{l,k}sin(x^{k}y^{l-k}).
\end{equation}
where $P_{i}(\cdot)$ and $P_{j}(\cdot)$ are Legendre polynomials. The highest degree of Legendre polynomials ($L_{l}$) and trigonometric functions ($L_{t}$) are set to $15$ and $2$, respectively. Weights $w$ are selected randomly from the range $[-20,20]$. In our experiment, the intensity of bias field is rescaled to range $[0.8,1.2]$ for low bias level and $[0.3,1.7]$ for high bias level. For every slice in BrainWeb dataset, we randomly create 20 different bias fields multiplied with the bias-free MR image to create bias-influenced data. Thus, for each sequence, we have 1810 bias-influenced MR images for training and another 1810 bias-influenced MR images for test. The simulated bias field, as the ground truth, are used to train ABCNet~\cite{ref_article11}.
\begin{table}[h]
\caption{Coefficient of variation (\%) comparisons among bias corrected methods on BrainWeb dataset. Brain tissues include CSF, GM and WM with low- and high-bias levels in T1 and T2 sequences. The metrics are indicated by mean and std values. The best three results are indicated by \textcolor{red}{red}, \textcolor{cyan}{cyan} and \textcolor{blue}{blue}, respectively.}\label{tab1}
\centering
\begin{tabular}{cccccccc}
\hline
\multicolumn{2}{c}{\multirow{2}{*}{Method}} & \multicolumn{3}{c}{low level}                   & \multicolumn{3}{c}{high level}                   \\ \cline{3-8} 
\multicolumn{2}{c}{}                        & CSF            & GM             & WM            & CSF            & GM             & WM             \\ \hline
\multirow{2}{*}{Input}          & T1        & 23.24$\pm$1.85 & 11.39$\pm$1.45 & 7.38$\pm$1.78 & 31.70$\pm$4.88 & 23.99$\pm$6.20 & 21.11$\pm$7.35 \\ \cline{2-8} 
                                & T2        & 14.24$\pm$1.16 & 18.17$\pm$1.70 & 9.48$\pm$2.95 & 25.01$\pm$5.16 & 28.30$\pm$5.33 & 22.50$\pm$7.37 \\ \hline
\multirow{2}{*}{N4\cite{ref_article4}}             & T1        & \textcolor{blue}{22.74$\pm$1.79} & \textcolor{blue}{9.69$\pm$1.40}  & \textcolor{cyan}{5.16$\pm$1.33} & 27.00$\pm$3.79 & 16.90$\pm$5.38 & 13.87$\pm$6.09 \\ \cline{2-8} 
                                & T2        & \textcolor{blue}{12.71$\pm$0.93} & \textcolor{cyan}{17.41$\pm$1.65} & \textcolor{red}{7.70$\pm$2.66} & 18.12$\pm$4.27 & 21.99$\pm$4.15 & 14.92$\pm$6.25 \\ \hline
\multirow{2}{*}{MICO\cite{ref_article7}}           & T1        & 26.40$\pm$3.67 & 13.71$\pm$4.29 & 8.76$\pm$4.22 & 30.50$\pm$5.11 & 20.13$\pm$5.55 & 16.28$\pm$6.21 \\ \cline{2-8} 
                                & T2        & 12.95$\pm$1.01 & 17.99$\pm$1.63 & 8.92$\pm$2.82 & 19.52$\pm$4.30 & 24.45$\pm$4.51 & 18.67$\pm$6.88 \\ \hline
\multirow{2}{*}{BCFCM\cite{ref_article8}}            & T1        & 23.42$\pm$2.71 & 10.77$\pm$3.10 & 6.11$\pm$4.29 & \textcolor{blue}{24.49$\pm$3.05} & \textcolor{blue}{12.26$\pm$3.39} & \textcolor{blue}{7.99$\pm$4.28}  \\ \cline{2-8} 
                                & T2        & \textcolor{cyan}{12.27$\pm$2.37} & 18.36$\pm$1.90 & \textcolor{blue}{8.12$\pm$2.86} & \textcolor{cyan}{13.27$\pm$2.73} & \textcolor{cyan}{19.18$\pm$2.24} & \textcolor{red}{9.50$\pm$3.38}  \\ \hline
\multirow{2}{*}{ABCNet\cite{ref_article11}}         & T1        & \textcolor{cyan}{22.09$\pm$1.51} & \textcolor{cyan}{9.37$\pm$1.20}  & \textcolor{red}{4.49$\pm$0.72} & \textcolor{red}{22.02$\pm$1.59} & \textcolor{red}{9.60$\pm$1.30}  & \textcolor{red}{4.77$\pm$0.83}  \\ \cline{2-8} 
                                & T2        & 13.41$\pm$1.04 & \textcolor{red}{17.30$\pm$1.69} & 8.17$\pm$3.05 & \textcolor{blue}{14.43$\pm$1.71} & \textcolor{red}{18.40$\pm$2.21} & \textcolor{cyan}{10.17$\pm$3.71} \\ \hline
\multirow{2}{*}{Ours}           & T1        & \textcolor{red}{21.90$\pm$1.79} & \textcolor{red}{8.97$\pm$1.46}  & \textcolor{blue}{6.01$\pm$3.53} & \textcolor{cyan}{23.83$\pm$2.56} & \textcolor{cyan}{11.97$\pm$2.89} & \textcolor{cyan}{7.03$\pm$2.63}  \\ \cline{2-8} 
                                & T2        & \textcolor{red}{11.73$\pm$0.85} & \textcolor{blue}{17.50$\pm$1.83} & \textcolor{cyan}{8.05$\pm$3.18} & \textcolor{red}{13.01$\pm$1.46} & \textcolor{blue}{19.24$\pm$2.27} & \textcolor{blue}{10.38$\pm$4.36} \\ \hline
\end{tabular}
\end{table}
\begin{table}[h]
\caption{Similarity between bias-free image and bias-corrected image on the BrainWeb dataset. The metrics include SSIM and PSNR indicated by mean and std values. The best three results are indicated by \textcolor{red}{red}, \textcolor{cyan}{cyan} and \textcolor{blue}{blue}, respectively.}\label{tab2}
\begin{tabular}{ccccccccc}
\hline
\multicolumn{3}{c}{}                                                                               & Input          & N4\cite{ref_article4}            & MICO          \cite{ref_article7} & BCFCM\cite{ref_article8}            & ABCNet        \cite{ref_article11} & Ours           \\ \hline
\multirow{4}{*}{T1} & \multirow{2}{*}{\begin{tabular}[c]{@{}c@{}}low \\ level\end{tabular}} & SSIM & 0.98$\pm$0.02  & \textcolor{cyan}{0.98$\pm$0.03}  & 0.92$\pm$0.05  & 0.67$\pm$0.07  & \textcolor{red}{0.99$\pm$0.03}  & \textcolor{blue}{0.97$\pm$0.04}  \\ \cline{3-9} 
                    &                                                                       & PSNR & 32.73$\pm$4.66 & \textcolor{cyan}{32.16$\pm$6.88} & 23.30$\pm$3.50 & 15.86$\pm$2.51 & \textcolor{red}{33.57$\pm$6.36} & \textcolor{blue}{30.42$\pm$6.88} \\ \cline{2-9} 
                    & \multirow{2}{*}{\begin{tabular}[c]{@{}c@{}}high\\ level\end{tabular}} & SSIM & 0.92$\pm$0.04  & \textcolor{blue}{0.94$\pm$0.04}  & 0.83$\pm$0.08  & 0.64$\pm$0.08  & \textcolor{red}{0.97$\pm$0.04}  & \textcolor{cyan}{0.95$\pm$0.04}  \\ \cline{3-9} 
                    &                                                                       & PSNR & 23.06$\pm$3.49 & \textcolor{blue}{24.45$\pm$4.03} & 19.42$\pm$3.33 & 15.47$\pm$2.61 & \textcolor{red}{27.74$\pm$0.04} & \textcolor{cyan}{26.40$\pm$3.85} \\ \hline
\multirow{4}{*}{T2} & \multirow{2}{*}{\begin{tabular}[c]{@{}c@{}}low\\ level\end{tabular}}  & SSIM & 0.98$\pm$0.01  & \textcolor{cyan}{0.98$\pm$0.01}  & 0.90$\pm$0.04  & 0.65$\pm$0.06  & \textcolor{red}{0.99$\pm$0.06}  & \textcolor{blue}{0.96$\pm$0.04}  \\ \cline{3-9} 
                    &                                                                       & PSNR & 30.39$\pm$3.88 & \textcolor{blue}{30.45$\pm$4.71} & 21.87$\pm$3.33 & 14.23$\pm$1.31 & \textcolor{cyan}{31.10$\pm$1.44} & \textcolor{red}{31.20$\pm$3.15} \\ \cline{2-9} 
                    & \multirow{2}{*}{\begin{tabular}[c]{@{}c@{}}high\\ level\end{tabular}} & SSIM & 0.89$\pm$0.04  & \textcolor{blue}{0.91$\pm$0.04}  & 0.74$\pm$0.09  & 0.61$\pm$0.06  & \textcolor{red}{0.98$\pm$0.01}  & \textcolor{cyan}{0.94$\pm$0.03}  \\ \cline{3-9} 
                    &                                                                       & PSNR & 20.50$\pm$3.32 & \textcolor{blue}{21.10$\pm$3.92} & 16.53$\pm$3.11 & 13.64$\pm$1.31 & \textcolor{red}{28.34$\pm$2.00} & \textcolor{cyan}{26.44$\pm$3.20} \\ \hline
\end{tabular}
\end{table}
\subsubsection{Experimental results}
\begin{figure}
\includegraphics[width=9.1cm]{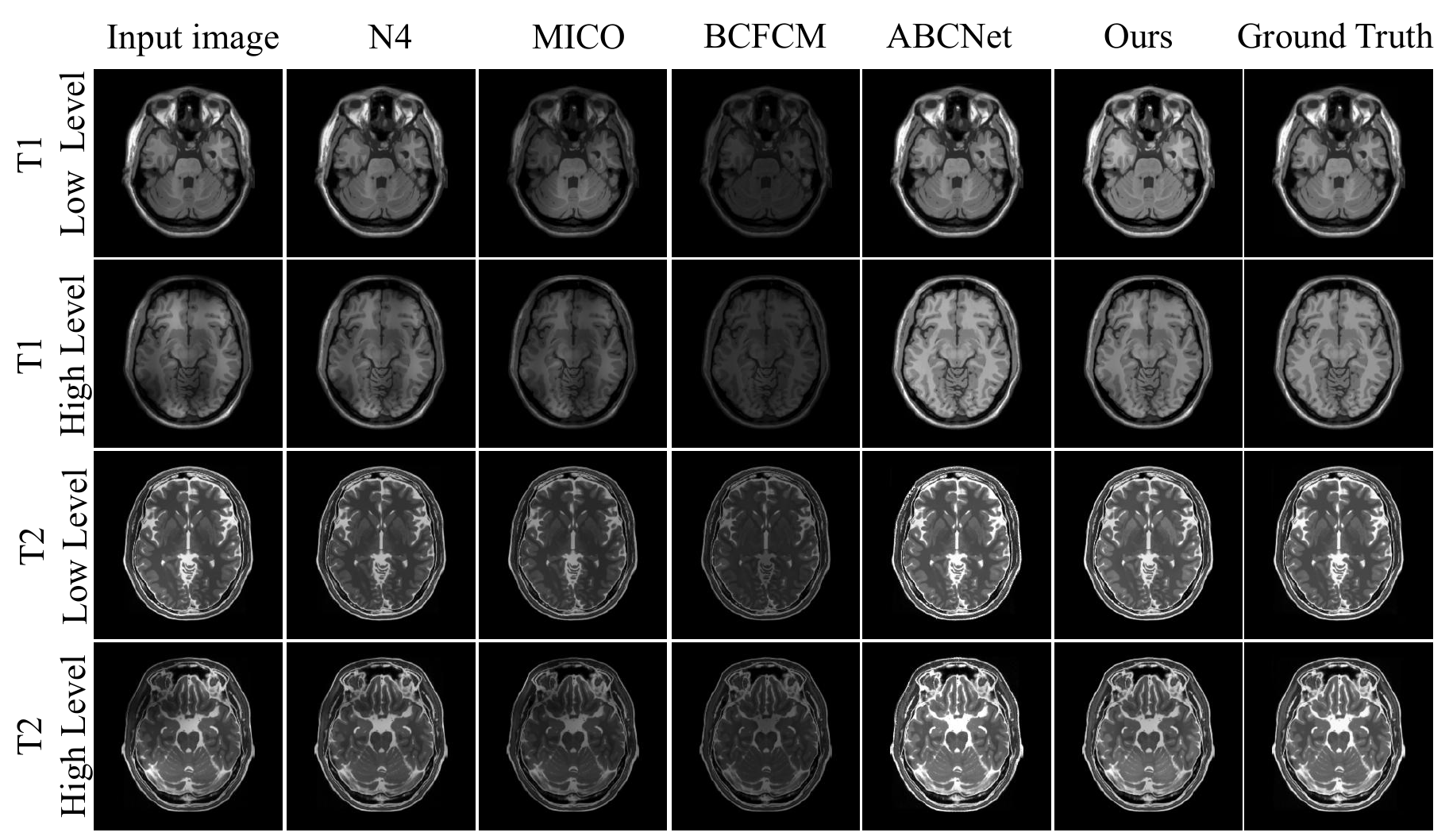}
\centering
\caption{Qualitative results of bias field correction on MR image of T1 and T2 modalities with low- and high- bias levels, respectively, where ground truth is the MR image without bias field.} \label{fig3}
\end{figure}

In our experiment, we compared our method to conventional methods N4~\cite{ref_article4}, MICO~\cite{ref_article7}, bias-corrected FCM (BCFCM)~\cite{ref_article8} and deep learning-based method ABCNet~\cite{ref_article11}. We also computed the metrics on input data for comparison. As the BCFCM method neglects the smoothness of bias field, we improved it by Eq.~\ref{bias}. 
The experimental results are demonstrated in Table~\ref{tab1} and Table~\ref{tab2}. We found that the smoothly varied bias field with low intensity has little influence on image structure. But for high-level bias field, it not only brings intensity inhomogeneity, but also degrades the structure of MR image. N4 method performs well in low bias level correction tasks, but the accuracy drops drastically when dealing with high bias level images. BCFCM restores images with low intensity variance. However, it is achieved by discarding important details and finally casues blurred, low-light results. MICO selects low-order function to simulate the bias field with less computational burden. But the real bias field is complex, the accuracy of simulation is limited and it may lead to a worse result. ABCNet could remove the bias field from MR image and attain the best results in most conditions. Our proposed method is on par with conventional methods and supervised methods. \textbf{For MR image with a high bias level, our method outperforms the N4 method in both intensity variance and similarity metrics.} Compared with the BCFCM, our method \textbf{restricts the range of bias fields in the output layer to protect the structure of image and adopts a learning strategy accompanied by momentum for escaping local extrema so as to obtain a better corrected result}. Different from ABCNet, which is supervised on both segmentation labels and bias field labels, our method is \textbf{only trained on the input image without any labels.} We also illustrate the corrected results in Fig.~\ref{fig3} for visual comparison. It can be seen that our method removes the bias field and preserves the structure details to obtain comparable results.
Different from the simulated dataset, the real datasets lack ground truth on the bias field for supervised learning, because of which ABCNet is inapplicable to the OAI-ZIB dataset. Thus, we compared our method to other conventional bias corrected methods, which are shown in Table~\ref{tab3}. We also made the experiment on knee segmentation based on U-Net. We found that the average Dice(\%) obtained an improvement from $97.91\pm2.42$ to $98.06\pm1.94$ with the bias correction of our method on the dataset. 
\begin{table}[h]
\caption{Coefficient of variation (\%) comparisons among bias correct methods on the OAI-ZIB dataset. The metrics are indicated by mean and std values. The best three results are indicated by \textcolor{red}{red}, \textcolor{cyan}{cyan} and \textcolor{blue}{blue}, respectively.}\label{tab3}
\centering
\begin{tabular}{|c|c|c|c|c|}
\hline
      & Femoral Bone   & Femoral Cartilage & Tibial Bone    & Tibial Cartilage \\ \hline
Input & 28.76$\pm$8.23 & 28.96$\pm$7.58    & 28.44$\pm$5.94 & 26.99$\pm$8.55   \\ \hline
N4\cite{ref_article4}    & \textcolor{cyan}{27.33$\pm$7.76} & 30.10$\pm$7.31    &\textcolor{cyan}{26.71$\pm$6.07} & 27.63$\pm$8.54   \\ \hline
MICO\cite{ref_article7}  & 28.68$\pm$8.52 & \textcolor{blue}{29.01$\pm$7.57}    & \textcolor{blue}{28.17$\pm$6.32} & \textcolor{blue}{27.07$\pm$8.49}   \\ \hline
BCFCM\cite{ref_article8}   & \textcolor{blue}{28.42$\pm$3.80} & \textcolor{cyan}{26.92$\pm$7.57}    & 28.74$\pm$3.58 & \textcolor{cyan}{25.35$\pm$8.14}   \\ \hline
Ours  & \textcolor{red}{27.08$\pm$4.62} & \textcolor{red}{24.16$\pm$7.98}    & \textcolor{red}{26.16$\pm$3.75} & \textcolor{red}{21.12$\pm$9.53}   \\ \hline
\end{tabular}
\end{table}
\section{Conclusion}
As segmentation map and bias field are highly associated, we proposed novel unsupervised bias decomposition networks, composed of a segmentation network and a bias estimation network, for bias correction of MR images. Besides, we designed reconstructed loss  functions for alternate training, based on which the smooth and multipilicative properties of bias field and intra-class consistency of segmentation map are ensured to remove the bias field and preserve the structure of MR image simultaneously. Experiments established on synthesized and clinical datasets show the effectiveness of our method on bias field correction tasks.
\subsubsection{Acknowledgements} ***.

\end{document}